# Kramer doublets, phonons, crystal-field excitations and their coupling in Nd$_2$ZnIrO$_6$

Birender Singh[1#], M. Vogl[2], S. Wurmehl[2,3], S. Aswartham[2], B. Büchner[2,3] and Pradeep Kumar[1*]

[1]*School of Basic Sciences, Indian Institute of Technology Mandi, Mandi-175005, India*
[2]*Leibniz-Institute for Solid State and Materials Research, IFW-Dresden, 01069 Dresden, Germany*
[3]*Institute of Solid-State Physics, TU Dresden, 01069 Dresden, Germany*

**Abstract:**

We report comprehensive Raman-scattering measurements on a single crystal of double-perovskite Nd$_2$ZnIrO$_6$ in temperature range of 4-330 K, and spanning a broad spectral range from 20 cm$^{-1}$ to 5500 cm$^{-1}$. The paper focuses on lattice vibrations and electronic transitions involving Kramer's doublets of the rare-earth Nd$^{3+}$ ion with local *C$_1$* site symmetry. Temperature evolution of these quasi-particle excitations have allowed us to ascertain the intricate coupling between lattice and electronic degrees of freedom in Nd$_2$ZnIrO$_6$. Strong coupling between phonons and crystal-field excitation is observed via renormalization of the self-energy parameter of the phonons i.e. peak frequency and line-width. The phonon frequency shows abrupt hardening and line-width narrowing below ~ 100 K for the majority of the observed first-order phonons. We observed splitting of the lowest Kramer's doublets of ground state ($^4I_{9/2}$) multiplets i.e. lifting of the Kramer's degeneracy, prominently at low-temperature (below ~ 100 K), attributed to the Nd-Nd/Ir exchange interactions and the intricate coupling with the lattice degrees of freedom. The observed splitting is of the order of ~2-3 meV and is consistent with the estimated value. We also observed a large number of high-energy modes, 46 in total, attributed to the intra-configurational transitions between 4f$^3$ levels of Nd$^{3+}$ coupled to the phonons reflected in their anomalous temperature evolution.

*#email id: birender.physics@gmail.com, *email id: pkumar@iitmandi.ac.in*

# Introduction

Physics of correlated electron systems associated with 5$d$ transition-metal oxides, especially Ir-oxides, has drawn considerable research interest in recent years owing to the possible formation of quantum-spin-liquids, Mott-insulators, unconventional superconductors and Weyl-semimetals [1-10]. Iridates provide an interesting interplay between strong spin-orbit coupling and electron correlation due to their comparable strength along with exotic quasi-particle excitations such as orbitons [11]. Double perovskite iridium-based materials of $A_2B$IrO$_6$ crystal structure are interesting as well as challenging due to the freedom of tuning the quantum magnetic ground state with the choice of different magnetic and non-magnetic ions on $A$ and $B$ crystallographic sites. So far, a majority of the studies have focused on tuning the magnetic exchange interactions by replacing $B$-sites with magnetic (i.e. Mn, Fe, Co, Ni and Cu) and non-magnetic (i.e. Y, Mg and Zn) elements [12-20]. The evolution of exotic quantum magnetic properties with the choice of $A$-site element in these materials is not extensively explored [21-22]. The substitution on $A$-site with different lanthanide rare-earth elements may provide another degree of freedom to give insight into the complex ground state in these 5$d$-materials. This replacement may cause substantial local structural changes (i.e. change of bond distance Ir-O, bond angle O-Ir-O, and rotation and/or tilting of IrO$_6$ octahedra) within the unit cell, and may result in significant modulation of the strength of exchange interactions between nearest-neighbors. Depending on the ionic-radius of $A$-site ions, different quantum magnetic spin interactions such as Heisenberg and Kitaev type may be realized in double pervoskite iridates [23-25]. Another important aspect of the presence of rare-earth ion in these systems is the crystal field excitations (CFE) originating from the splitting of rare-earth ion multiplets with specific local site symmetry owing to the surrounding ions static field. Probing these electronic excitations are very important because CFE as a function of temperature may provide crucial information about the nature of underlying ground state ordering and also provide the avenues to probe the coupling with lattice degrees of freedom [26-29].

The magnetic phase of $Nd_2ZnIrO_6$ is rather mysterious. It shows an antiferromagnetic ordering at ~ 17 K attributed to the ordering of $Nd^{3+}$ ions with a small contribution from Ir sub-lattice, along with a broad transition at 8 K. However, under weak magnetic field, a complex magnetic phase is reported with multiple magnetic transitions [22] attributed to the interplay between Nd and Ir magnetic sub-lattices. $Nd^{3+}$ ion here inhabits an eight-fold coordinated *A*-site with $C_1$ local site symmetry that shows there is no symmetry element except identity. The sensitivity of the $Nd^{3+}$ crystal-field levels to both local electric and magnetic fields makes Raman scattering by these CF levels a valuable probe for exploring the possible underlying interactions responsible for their ground state. Measurements of CFE are also important to understand various thermo-dynamical properties and accurately determine the generalized phonon density of states. Their measurements via Raman is complementary to complex neutron-scattering measurements, where one requires a big single-crystal. Here, we present a detailed study of a single-crystal of double-perovskite $Nd_2ZnIrO_6$ focusing on the phonons, CFE, and the coupling between them. We note that not much attention has been paid for understanding the role of phonons and the coupling with other quasi-particle excitations in these Ir-based double-perovskite systems. We have performed a detailed lattice-dynamic study of $Nd_2ZnIrO_6$ in the temperature range of 4 K to 330 K, and polarization-dependent Raman-scattering measurements along with the zone-centered phonon mode frequency calculations using first-principle based density functional theory (DFT). Our measurements reveal strong coupling between phonons and CFE, and evidence the lifting of Kramer's degeneracy of $Nd^{3+}$ ground state multiplets at low-temperature attributed to the Nd-Nd and Nd-Ir exchange interactions.
3

**Experimental and computational details**

The inelastic light (Raman) scattering measurements were performed on a single-crystal of $Nd_2ZnIrO_6$ in back scattering configuration via Raman spectrometer (LabRAM HR- Evolution). We have employed 532 nm (2.33 eV) and 633 nm (1.96 eV) lasers as an excitation probe for spectral excitation. The laser was focused on the sample surface using 50x long working distance objective lens. A Peltier-cooled charge-coupled device (CCD) detector was used to collect the scattered light. Laser power was limited ($\leq 1mW$) to avoid any local heating effects. Temperature-dependent measurements were performed with a closed-cycle He-flow cryostat (Montana Instruments), where sample is mounted on top of the cold finger inside a chamber, which was evacuated to $1.0 \times 10^{-6}$ Torr, in the temperature range of 4 K to 330 K with $\pm$ 0.1 K or better temperature accuracy.

Density functional theory-based calculations were done to have insight about phonon dynamics using plane-wave approach implemented in QUANTUM ESPRESSO [30]. The fully-relativistic Perdew-Burke-Ernzerhof (PBE) was chosen as an exchange correlation functional within generalized gradient approximation to carry out these calculations. A plane wave cut-off energy and charge density cut-off was set to 60 Ry and 280 Ry, respectively. The numerical integration over the Brillouin-zone was done with a 4 x 4 x 4 k-point sampling mesh in the Monkhorst Pack [31]. Dynamical matrix and zone-center phonon frequencies were calculated using density functional perturbation theory, taking spin-orbit coupling without spin-polarization into account [32]. All the calculations were performed with experimental lattice parameters reported in ref. 22.

**Results and Discussion**

    **A. Raman-scattering and Zone-centered calculated phonon frequencies**

$Nd_2ZnIrO_6$ crystallizes in a monoclinic double-perovskite structure belonging to the *P2$_1$/n* space group (No. 14) and *C$_{2h}$* point group [22]. Within this structure the factor-group analysis predicts a total of 60 modes in the irreducible representation, out of which 24 are Raman active and 36 are infrared active (see Table-I for details). The Raman spectra of $Nd_2ZnIrO_6$ single-crystals were excited with two different incident photon energies 633 nm (1.96 eV) and 532 nm (2.33 eV). Figure 1 shows the Raman spectra of $Nd_2ZnIrO_6$ excited with 633 nm laser at 4 K, while the insets shows its temperature evolution and comparison with 532 nm at 4 K and 330 K. We note that the Raman spectra when excited with 532 nm exhibits extra features in the spectral range of 50-800 $cm^{-1}$ and 1700-5500 $cm^{-1}$, labeled as R1-R7 (see inset a2 of Fig. 1) and Q1-Q46 (see Figure 6 (a and b)), respectively. A detailed discussion about these modes is discussed in later sections. First, we will focus on the first-order phonon modes shown in Figure 1. We notice a total of 23 phonon modes in the spectrum excited with 633 nm laser, labeled as P1-P23. The data is fitted with a sum of Lorentzian functions to obtain the spectral parameters such as phonon mode frequency ($\omega$), line-width ($\Gamma$) and the integrated intensity.

In order to have insight on the specific vibration patterns of the observed phonon modes, we carried out zone-centered lattice dynamics calculations using density functional perturbation theory. The calculated mode frequencies together with atomic displacements (vibrations) associated with particular phonon modes of $Nd_2ZnIrO_6$ are listed in Table-II. The calculated phonon modes frequencies are found to be in very good agreement with the experimentally observed values. Figure 2 illustrate the schematic representations of the calculated eigen-vectors of the corresponding phonon modes. The evaluation of these eigen-vectors reveals that the low-frequency phonon modes P1-P3, P5 are mainly attributed to the atomic displacement of Nd-atoms, while P4 is composed of the displacement of Nd/Zn/Ir. The high-frequency phonon modes P6 to P23 are corresponds to the stretching and bending of the Zn/Ir-O bonds in the $Zn/IrO_6$ octahedra.

The phonon modes assignment is done in accordance to our density functional theory calculations and polarization-dependent Raman measurements.

**B. Temperature and polarization-dependence of the first-order phonon modes**

Figure 3 shows the temperature-dependence of the frequency and line-width of the first-order phonon modes of $Nd_2ZnIrO_6$ excited with 633 nm laser. The following important observation can be made: (i) Low-energy phonon mode P1 (~ 100 cm$^{-1}$) exhibit phonon softening below ~ 100 K, however no anomaly is observed in the line-width of this mode in the entire temperature range. (ii) Interestingly, the phonon modes P3-P4, P7-P8, P10-P11, P15 and P21-P22 show significant frequency hardening with a decrease in temperature below ~ 100 K down to the lowest recorded temperature, while P20 and P23 phonon modes display a slight softening. Additionally, a clear change in the line-widths of all these phonon modes is observed around ~ 100 K, i.e. anomalous line-narrowing below this temperature. Similar temperature-dependence is also observed for the frequency and line-width of phonon modes excited with 532 nm laser (not shown here). We note that the anomalies are more pronounced for the low-frequency phonon modes. The observed anomaly in the phonon modes frequencies and line-widths at low-temperature clearly indicate strong interaction of the phonons with other degrees of freedom. Long-range magnetic ordering in this system is reported at ~ 17 K [22], however the phonon anomalies are observed at temperature as high as ~ 100 K, well above the magnetic transition temperature. This observation suggests its origin from other degrees of freedom in the observed phonons anomalies than magnetic ones, such as electronic degrees of freedom. Here, a coupling to crystal-field multiplets may be at work.

To understand the temperature-dependence of the first-order phonon modes, we have fitted the phonon mode frequency and line-width from ~ 100 K to 330 K with an anharmonic phonon-phonon interaction model including both three and four-phonons decay channels given as [33]:

$$\omega(T) = \omega_0 + A(1+\tfrac{2}{e^x-1}) + B(1+\tfrac{3}{e^y-1}+\tfrac{3}{(e^y-1)^2}) \text{ and,} \qquad (1)$$

$$\Gamma(T) = \Gamma_0 + C(1+\tfrac{2}{e^x-1}) + D(1+\tfrac{3}{e^y-1}+\tfrac{3}{(e^y-1)^2}) \qquad (2)$$

where $\omega_0$ and $\Gamma_0$ are the mode frequency and line-width at T = 0 K, respectively, $x = \frac{\hbar\omega_0}{2k_BT}$; $y = \frac{\hbar\omega_0}{3k_BT}$ and A, B, C and D are constant. The fitted parameters are summarized in Table-II. The expected temperature variation of mode frequency and line-width due to anharmonicity is shown as solid red and blue lines (see Fig. 3). Within the anharmonic interaction picture phonon modes are expected to become sharper and mode frequency gradually increase with decreasing temperature. Notably, at low-temperature below ~ 100 K the phonon mode self-energy parameters, i.e. mode frequencies and line-widths, exhibit significant deviation from the curve estimated by anharmonic phonon-phonon interaction model. Majority of the observed phonon mode frequencies harden anomalously below ~ 100 K, except mode P1, P20 and P23. Interestingly, the magnitude of the phonon frequency hardening is observed to be more pronounced for the low-frequency phonon modes (below 300 cm$^{-1}$). In general, the magnitude of anomalous modulation of phonon frequencies are usually attributed to the interaction strength of a particular phonon mode with, e.g., spin degrees of freedom. Hence, in a long-range magnetic phase the phonon renormalization mainly occurs due to the strong coupling of lattice with spin degrees of freedom. The effect of magnetic degrees of freedom is expected to be minimal if not zero in this case because $T_N$ (~ 17 K) is much lower than the temperature (~ 100 K) where anomalies are observed. In this scenario, the pronounced self-energy renormalization of the phonon modes implies a strong coupling of lattice with electronic degrees of freedom via crystal-field excitations associated 4f-levels of Nd$^{3+}$ ion. Indeed, we found signature of strong coupling of these two degrees of freedom where CFE excitations, which are also in the same energy range and have similar symmetry, arising from ground state ($^4I_{9/2}$) multiplets of Nd$^{3+}$ ion at low-temperature and led to the strong renormalization of the phonon self-energy parameters.

To decipher the symmetry of the phonon modes, we also performed polarization-dependent measurements. Polarization-dependent measurements using Raman spectroscopy can be done using equivalent configurations based on the polarization of incident or scattered light or rotation of the sample. In our polarization-dependent study, polarization of the incident laser light is varied i.e. incident laser polarization is rotated using a polarizer with an increment of 10º from 0º to 360º, while the direction of the scattered light polarization is fixed using an analyzer. Figure 4 (a and b) illustrate the polarization-dependence of the Raman spectra of Nd$_2$ZnIrO$_6$ at different angle of incident light polarization vector on the sample surface ranging from 0 - 90º in the spectral range of 50-330 cm$^{-1}$ and 330-800 cm$^{-1}$, respectively (inset of Fig. 4 (a and b) shows Raman spectra at different angle from 90-180º). Angular dependence of the intensity of the prominent phonon modes is shown in Figure 4 (d). Intensity of these prominent first-order phonon modes shows two-fold symmetric nature i.e. having maximum intensity at $0^0/90^0$ and $180^0/270^0$. One may understand the observed variation in the intensity as a function of the polar-angle within the semi-classical approach. As the incident and scattered polarized light lies in the *ab* plane, the unit vector associated with the incident ($\hat{e}_i$) and scattered ($\hat{e}_s$) light polarization may be expressed in the decomposed form as $(\cos(\theta+\theta_o) \ \sin(\theta+\theta_o) \ 0)$ and $(\cos(\theta_o) \ \sin(\theta_o) \ 0)$, respectively, for the case of propagation of linear-polarized light in the z-direction (see right panel of Fig. 4 (c)). Within the semi-classical approach, the Raman-scattering cross-section is given as $I_{Raman} \propto |\hat{e}_s^t.R.\hat{e}_i|^2$, where t denotes the transpose of $\hat{e}_s$ and R (see Table-I) is the Raman tensor [34-36]. Using above expression for the intensity, angular dependence of Raman intensity for the mode *A$_g$* and *B$_g$* are given as $I_{A_g} = | a.\cos(\theta_o).(\cos(\theta+\theta_o)) + b.\sin(\theta_o).\sin(\theta+\theta_o)) |^2$ and $I_{B_g} = | e.\cos(\theta_o).(\sin(\theta+\theta_o)) + e.\sin(\theta_o).\cos(\theta+\theta_o)) |^2$. Here $\theta_o$ is an arbitrary angle from *a*-axis and is constant. Therefore, without the loss generality it may be chosen to be zero, giving rise to the expression for the Raman intensity as $I_{A_g} = |a|^2 \cos^2(\theta)$ and $I_{B_g} = |e|^2 \sin^2(\theta)$. From these intensity expressions it is clear that these two types of phonon modes are out of phase with each other i.e. the angular dependence of $A_g$ mode intensity decreases to zero when incident and scattered polarization vectors are perpendicular to each other. On the other hand, the intensity of $B_g$ phonon mode shows maxima in this configuration. The solid lines in Fig. 4 (d) represent the fitted curves using the above expression for the intensity of the phonon modes. The fitted curves are in very good agreement with the experimental angle variation of the mode intensities. Using the polarization-dependence of the phonon modes and our DFT based calculations, we have assigned the symmetry of the observed phonon modes (see Table-II). Symmetry of the phonon modes extracted from our polarization-dependent measurement is mapped nicely with the mode symmetries calculated using the first principle DFT based calculations.

### C. Crystal-field excitations and coupling with phonons

In addition to the phonon modes several additional modes are observed in the spectrum excited with 532 nm (2.33 eV) laser. First, we will discuss the modes observed in the low-frequency range below ~ 500 cm$^{-1}$, labeled as R1-R7 (see Fig. 5(c) and inset a2 of Fig. 1), which appear in low-temperature regime below ~ 150-200 K. We have assigned these modes to the crystal-field excitations within the ground state multiplets of Nd$^{3+}$ ($^4I_{9/2}$, ground state has five Kramer's doublets, see inset of Fig. 5(c) for their energy levels [37-38]). As the frequencies of Raman active modes are independent of the incident photon energy, and as result appear at the same position in Raman-shift even with different energy of the excitation source; also, Raman active modes appear at the same position in both stokes and anti-stokes side of the spectrum. However, the emission or absorption peak arising due to transitions between different electronic levels will be shifted equal to the separation between the two incident laser energies, whereas their absolute-position in energy must be preserved. These low-energy modes are visible in both stokes and anti-stokes side of the spectrum with 532 nm laser at the same position suggesting Raman active nature of these excitations.

Figure 5 (a, b) shows the temperature-dependence of the frequency, line-width and intensity of the prominent modes i.e. R2, R4 and R7. Shift in their frequency is quite significant and is as high as ~ 3% (for R4) within a short interval of temperature suggesting intricate coupling with the phonon modes of similar energy and symmetry. The integrated intensity of these modes also increases at a very fast rate with decreasing temperature (see left side of Fig. 5(b)), which is typical for Raman-scattering associated with crystal-field transitions (because the ground state population increases with decreasing temperature) and shows temperature evolution opposite to those of phonons (see right side of Fig. 5(b) showing intensity for modes P1, P3 and P4). We also did the polarization-dependence of these modes at low-temperature (see Fig. 5 (d) and inset of Fig. 5(c)). The polarization-dependence of these modes reflect their quasi-isotropic nature and are quite different from the observed phonon modes polarization behavior, confirming the non-phononic nature of these modes. From the symmetry analysis, CF levels of the $Nd^{3+}$ in ground state (i.e. $^4I_{9/2}$) multiplets split into five Kramer's doublets. Since $Nd^{3+}$ has a $C_1$ local site symmetry, all the levels have *A* symmetry and transition between them are expected to be active in all polarizations [39]. Therefore, the observation of quasi-isotropic nature of the polarization-dependent behavior of these CF modes is in-line as predicted by the symmetry analysis. We note that these lower-energy crystal-field modes appear only at low temperature (below ~ 200 K). The Raman cross-section for the CFE is often weak and these have been observed generally with the strong coupling with nearby Raman active phonon modes via strong electron-phonon interaction. Two quasi-particle excitations of similar energy and symmetry are expected to couple strongly [39], therefore a phonon mode and CFE of same symmetry and similar energy are also expected to couple strongly. As discussed in the previous section above, renormalization of the phonon modes at low-temperature (~ 100 K) is very strong (see Fig. 3) and interestingly renormalization effect are more pronounced for low-energy phonon modes (e.g. P1, P3, P8, P10 and P11; all have $A_g$ symmetry). Energy range of these phonon modes (maximum ~ 400 cm$^{-1}$) is similar to those of CFE (maximum ~ 450 cm$^{-1}$) and both these excitations are of similar symmetry. Therefore, our observation of renormalization of the phonon modes at low-temperature is attributed to the strong coupling with these crystal-field excitations.

Another interesting observation is that the number of possible transitions within the ground state multiplets at low-temperature are expected to be only four as shown in the inset of Figure 5 (c), however seven modes are observed (R1-R7) i.e. three doublet (R1-R2, R3-R4, R5-R6) and mode R7. At low-temperature we observed four strong modes i.e. R2, R4, R6 and R7 and with further decrease in temperature three weak shoulder modes namely R1, R3 and R5 also started gaining spectral weight. Frequency difference between these doublets is ~ 2-3 meV (~ 20-25 cm$^{-1}$) and it increases with decrease in temperature (see inset in top panel of Fig.5(a)). The observed weak modes i.e. R1, R3 and R5 may arise from the splitting of the ground state Kramer's doublet by the Nd-Nd and Nd-Ir exchange interactions. Interestingly, with increasing temperature the overall intensity of the higher-frequency doublet components ( i.e. R2, R4 and R6 ) decreases relative to the lower-frequency ones (i.e. R1, R3 and R5) (see Fig. 5 (c)). Also, discontinuities in the intensity follow the long-range magnetic ordering (see inset in top and middle panel in Fig.5 (b)) associated with the two interpenetrating AFM sub-lattices associated with the Nd and Ir moments ordering. This discontinuity is also visible in the doublet splitting (see inset in the top panel of Fig. 5 (a)). Quantitatively, the origin of these discontinuities may be understood using the fact that splitting of the Kramer's doublets is due to Nd-Nd and Nd-Ir exchange interactions, as these exchange interactions are also responsible for the magnetic ordering, along with a contribution from the dipolar magnetic field due to Ir ion on the Nd-atom. The observed magnetic transition at ~ 17 K roughly corresponds to the exchange interaction strength of ~ 2 meV, therefore the splitting between the ground state Kramer's doublet owing to the exchange interactions is expected to be of this order. Our observed splitting of ~ 25 cm$^{-1}$ is also close to the value as suggested by the exchange interaction strength. Further the splitting between ground state Kramer's doublet may also be estimated using the intensity ratio of the doublets and fitting the ratio (see inset in bottom panel of Fig. 5 (b)) using the functional form as [40]: $I = I_0 e^{\Delta/k_b T}$, where $I_0$ is temperature independent intensity ratio and $\Delta$ is the splitting of the ground state Kramer's doublet. Solid lines in the inset of bottom panel in Figure 5 (b) shows the good agreement between data and fit. The extracted value of the splitting of the ground state Kramer's doublet ($\Delta$) is ~ 30 cm$^{-1}$ and is consistent with our observation.

**D. Intra-configurational transitions and interaction with phonons**

We observed a large number of sharp and strong modes, labeled as Q1-Q46, in the high-energy range spanning from 1700 to 5500 cm$^{-1}$, shown in Figure 6 (a, b). The insets show spectra in the absolute energy scale from 17100 to 13300 cm$^{-1}$. The origin of these sharp modes may be understood as follows. When the Nd$^{3+}$ is placed in $C_1$ local site symmetry as in the present case, the ground ($^4I_{9/2}$), first excited ($^4I_{11/2}$) and second excited ($^4I_{13/2}$) state multiplets split into five, six and seven Kramer's doublets, respectively, and so on. In fact, because of the lowest local symmetry at $Nd^{3+}$ ions i.e. $C_1$, the degeneracy of the free-ion levels must be completely lifted so that (2J + 1) levels are expected for each J term, which further reduces to (2J + 1)/2 levels as constrained by the Kramer's theorem. The doublets of $^4G_{5/2} +^2 G_{7/2}$, $^4F_{9/2}$ and $^4F_{7/2} +^4 S_{3/2}$ are at ~ 17000 cm$^{-1}$, 14700 cm$^{-1}$, and 13800 cm$^{-1}$, respectively, [37-38] (see Fig. 6(c) for schematic of crystal-field splitting of these levels). Based on the comparative energy with different CF levels, modes from Q1-Q20 may be assigned to the transitions from higher lying doublets of ($^4G_{5/2} +^2 G_{7/2}$) levels to the different doublets of the ground state ($^4I_{9/2}$). Similarly, modes Q21-Q38, and Q39-Q46 may be assigned to transitions from $^4F_{9/2}$ and $^4F_{7/2} +^4 S_{3/2}$ to the doublets of ground state ($^4I_{9/2}$), respectively. Figure 6 (a, b) shows the modes attributed to the transitions from higher levels to the ground state multiplets, say from $^4G_{5/2} /^2 G_{7/2}$ to the $^4I_{9/2}$ in the vicinity of ~ 2000 cm$^{-1}$ (~ 16800 cm$^{-1}$ in absolute scale). From this figure, we observed a continuous increase in the line-widths, say as one goes from mode Q6 to Q20; and from Q22 to Q38 (see Table-III for the list of prominent modes line-width with this increasing trend), attributed to the strong electron-phonon interaction. We know that a narrow line-width is associated with the electronic transitions from the lowest multiplet of any crystal-field level to the lowest level of other multiplets of CF level, and it increases for transitions to the higher multiplets of the same CF level [41]. The underlying mechanism for such a broadening of the line-width is due to the possible relaxation to the available lower-energy levels with the spontaneous emission of phonons.

Figure 7 (a, b and c) shows the temperature-dependence of the prominent modes in the energy range of 13000-18000 cm$^{-1}$ i.e. Q2, Q5-Q7, Q10-Q11, Q15-Q16, Q22-Q23 and Q28-Q29. Following observations can be made: (i) The frequency of modes Q2, Q5-Q7, Q10-Q11, Q15-Q16 and Q22-Q23 are observed to exhibit significant blue-shift with decrease in temperature up to ~

100 K, on further decrease in temperature all these modes are red-shifted except mode Q10. (ii) Interestingly, the frequency of Q28 and Q29 decreases with decreasing temperature down to 4 K. (iii) The line-width of all the modes shows line-narrowing with decreasing temperature, the decrease in line-width is as large as 400-500%, for example see mode Q6, Q7, Q10, Q11. (iv) Intensity of all the observed modes show increase with decrease in temperature up to ~ 100 K, and on further decrease in temperature it decreases for mode Q2, Q5-Q7, Q10, Q22-Q23 and Q29. Intensity of Q11, Q15 is almost independent of temperature below ~ 100 K, however intensity of modes Q16 and Q28 keep increasing with decrease in temperature. Temperature evolution of these electronic transitions may be understood by using the fact that the renormalization of the frequencies and their line-width is intimately linked with the lattice degrees of freedom. Frequency shift is associated with the change in the phonon energy coupled with the transitions and the expression for the shift in energy for the electronic transition is given as [41]:

$$E_i(T) = E_{0i} + \alpha_i \left(\frac{T}{\theta_D}\right)^4 \int_0^{\theta_D/T} \frac{X^3}{e^X - 1} dX \qquad (3)$$

where, $E_{0i}$ is the energy at 0 K, $\alpha_i = \frac{3}{2} W_i \omega_D$, $W_i$ is the ion-phonon coupling constant, $\omega_D$ is the Debye phonon energy, and $\theta_D$ is the Debye temperature taken as ~ 950 K from our density of phonon calculations. We have fitted the frequency of the prominent mode using the above expression, see solid lines in Figure 7 (a, b) and the fitting parameters are given in Table-IV. Fitting is in good agreement in the temperature range of 330 K to ~ 100 K, below 100 K it starts diverging from the expected behavior. The majority of these modes show a normal temperature dependence of the frequencies, i.e. increase in energy with decreasing temperature, until ~ 100 K. The temperature evolution may be gauged from the fact that these electronic transitions are mediated by phonons, and they are expected to show similar temperature-dependence as that of phonons.

From the fitting, the ion-phonon interaction parameter ($W_i$) is negative for these modes (see Table-IV). On the other hand, the temperature-dependence of mode Q28 and Q29 is anomalous i.e. peak frequency decrease with decreasing temperature, also reflected in the positive value of $W_i$ for mode Q29. We note that such anomalous temperature evolution has also been observed for different systems [42-44].

**Summary and conclusion**

In conclusion, anomalous renormalization of the first-order phonon modes below ~ 100 K is attributed to the strong coupling between lattice degrees of freedom and crystal-field excitations. The phonon mode frequencies and their symmetries are estimated using first-principle density functional theory-based calculations, which are in very good agreement with our polarization dependent Raman measurements. We observed splitting of ground state Kramer's doublets attributed to the exchange interaction between the Nd-Nd/Nd-Ir, and the observed doublet splitting is in very good agreement with the estimated values. Our results shed crucial light on the role of *A*-site rare-earth element in these iridium based double-perovskite materials reflected via coupled lattice and electronic degrees of freedom, and believe that our results will prove to be an important step in understanding the exotic ground state of these systems by considering electronic and phononic degrees of freedom on equal footing. We hope that our detailed experimental studies reveling a large number of intra-configurational transitions between CF-levels of $Nd^{3+}$ at low-site symmetry also provide a good starting point to calculate the energy of all possible CF-levels in these important class of double-perovskite systems using theoretical models.

**Acknowledgment**

PK thanks the Department of Science and Technology, India, for the grant and IIT Mandi for the experimental facilities. The authors at Dresden thank Deutsche Forschungsgemeinschaft (DFG) for financial support via Grant No. DFG AS 523/4-1 (S.A.) and via project B01 of SFB 1143 (project-id 247310070).

**Table-I:** Wyckoff positions of the different atoms in the unit cell and irreducible representations of phonon modes of the monoclinic ($P2_1/n$; space group No. 14) double-perovskite Nd$_2$ZnIrO$_6$ at $\Gamma$ - point. R$_{Ag}$ and R$_{Bg}$ are the Raman tensors for $A_g$ and $B_g$ Raman active phonon modes, respectively. $\Gamma_{total}$, $\Gamma_{Raman}$ and $\Gamma_{infrared}$ represents the total normal modes, Raman and infrared active modes, respectively.

| $P2_1/n$; space group No. 14 | | | |
|---|---|---|---|
| Atom | Wyckoff site | $\Gamma$-point mode decomposition | Raman tensors: |
| Nd | 4e | $3A_g + 3A_u + 3B_g + 3B_u$ | $R_{A_g} = \begin{pmatrix} a & 0 & d \\ 0 & b & 0 \\ d & 0 & c \end{pmatrix}$ |
| Zn | 2d | $3A_u + 3B_u$ | |
| Ir | 2c | $3A_u + 3B_u$ | |
| O(1) | 4e | $3A_g + 3A_u + 3B_g + 3B_u$ | $R_{B_g} = \begin{pmatrix} 0 & e & 0 \\ e & 0 & f \\ 0 & f & 0 \end{pmatrix}$ |
| O(2) | 4e | $3A_g + 3A_u + 3B_g + 3B_u$ | |
| O(3) | 4e | $3A_g + 3A_u + 3B_g + 3B_u$ | |

Classification:

$\Gamma_{total} = 12A_g + 18A_u + 18B_u + 12B_g$, $\Gamma_{Raman} = 12A_g + 12B_g$ and $\Gamma_{infrared} = 18A_u + 18B_u$

**Table-II:** List of DFT based calculated phonon frequencies and the experimentally observed first-order phonon mode frequencies of Nd$_2$ZnIrO$_6$ at 4 K with 633 nm excitation laser and fitting parameters obtained via using equations as described in the text. Modes assignment is done in accordance to polarization-dependent study and lattice-dynamics calculations. Units are in cm$^{-1}$.

| Mode Assignment | Exp. ω (4 K) | DFT ω | Fitted Parameters | | | | | |
|---|---|---|---|---|---|---|---|---|
| | | | $\omega_0$ | A | B | $\Gamma_0$ | C | D |
| **P1** - $A_g$ (Nd) | 100.7 | 93.6 | 101.4 ± 0.2 | -0.01 ± 0.03 | -0.003 ± 0.001 | 3.2 ± 0.6 | 0.06 ± 0.02 | 0.003 ± 0.001 |
| **P2** - $B_g$ (Nd) | 137.1 | 124.0 | | | | | | |
| **P3** - $A_g$ (Nd) | 143.3 | 131.2 | 142.5 ± 0.2 | -0.16 ± 0.11 | -0.003 ± 0.002 | 2.8 ± 0.2 | 0.15 ± 0.08 | 0.005 ± 0.002 |
| **P4** - $B_g$ (Nd,Zn,Ir) | 163.9 | 131.9 | 163.8 ± 0.3 | -0.26 ± 0.13 | -0.009 ± 0.003 | 2.4 ± 0.3 | 0.26 ± 0.18 | 0.008 ± 0.005 |
| **P5** - $B_g$ (Nd) | 193.6 | 150.5 | | | | | | |
| **P6** - $B_g$ (O) | 211.6 | 207.0 | | | | | | |
| **P7** - $B_g$ (O) | 224.7 | 220.9 | 223.4 ± 0.6 | -1.18 ± 0.07 | -0.007 ± 0.004 | 7.2 ± 3.4 | 2.71 ± 1.13 | -0.004 ± 0.002 |
| **P8** - $A_g$ (O) | 235.2 | 234.1 | 234.0 ± 1.2 | -0.49 ± 0.05 | -0.017 ± 0.003 | 8.2 ± 1.8 | 0.27 ± 0.13 | 0.039 ± 0.012 |
| **P9** - $B_g$ (O) | 259.4 | 238.8 | | | | | | |
| **P10** - $A_g$(O) | 263.5 | 276.1 | 262.7 ± 0.5 | -0.46 ± 0.27 | -0.015 ± 0.001 | 4.0 ± 0.7 | 0.92 ± 0.41 | 0.023 ± 0.017 |
| **P11** - $A_g$ (O) | 294.2 | 279.3 | 294.7 ± 0.4 | -1.09 ± 0.34 | -0.02 ± 0.01 | 3.6 ± 0.7 | 1.18 ± 0.51 | 0.064 ± 0.032 |
| **P12** - $A_g$ (O) | 362.5 | 328.5 | | | | | | |
| **P13** - $B_g$ (O) | 380.7 | 357.0 | 382.4 ± 0.6 | -2.29 ± 0.59 | -0.03 ± 0.02 | 4.8 ± 2.3 | 1.99 ± 0.54 | 0.234 ± 0.146 |
| **P14** - $A_g$ (O) | 397.8 | 372.6 | | | | | | |
| **P15** - $B_g$ (O) | 414.4 | 388.8 | 415.2 ± 0.3 | -0.91 ± 0.31 | -0.07 ± 0.02 | 4.9 ± 1.0 | 1.67 ± 0.67 | 0.168 ± 0.071 |
| **P16** - $B_g$ (O) | 453.2 | 423.4 | | | | | | |
| **P17** - $B_g$ (O) | 471.7 | 441.4 | 471.7 ± 1.5 | -0.32 ± 0.11 | -0.24 ± 0.13 | 7.4 ± 3.1 | 3.53 ± 1.57 | 0.113 ± 0.021 |
| **P18** - $A_g$ (O) | 532.3 | 538.3 | | | | | | |
| **P19** - $B_g$ (O) | 567.3 | 541.6 | | | | | | |
| **P20** - $B_g$ (O) | 584.9 | 558.0 | 587.7 ± 0.6 | -2.32 ± 0.76 | -0.04 ± 0.01 | 6.0 ± 2.56 | 2.71 ± 0.91 | 0.515 ± 0.211 |
| **P21** - $A_g$ (O) | 597.0 | 582.1 | 601.1 ± 1.3 | -4.20 ± 1.38 | -0.03 ± 0.01 | 6.3 ± 2.93 | 4.62 ± 2.31 | -0.088 ± 0.037 |
| **P22** - $A_g$ (O) | 656.9 | 668.4 | 660.4 ± 1.4 | -3.56 ± 1.58 | -0.27 ± 0.19 | 7.63 ± 3.90 | 8.73 ± 3.34 | 0.378 ± 0.171 |
| **P23** - $A_g$ (O) | 673.5 | 721.7 | 673.5 ± 0.8 | 0.71 ± 0.18 | -0.33 ± 0.11 | 4.37 ± 2.61 | 5.64 ± 2.34 | 0.364 ± 0.113 |

**Table-III:** List of the frequency and line-width of the prominent intra-configurational transition modes at 75 K. Units are in cm$^{-1}$.

| $^4G_{5/2}$ + $^2G_{7/2}$ | | | $^4F_{9/2}$ | | | $^4F_{7/2}$ + $^4S_{3/2}$ | | |
|---|---|---|---|---|---|---|---|---|
| **Mode** | $\omega_{abs.}$ | $\Gamma$ | **Mode** | $\omega_{abs.}$ | $\Gamma$ | **Mode** | $\omega_{abs.}$ | $\Gamma$ |
| Q6  | 16,875 | 5.2  | Q22 | 14,990 | 9.8  | Q39 | 14,460 | 16.3  |
| Q7  | 16,848 | 10.6 | Q23 | 14,938 | 5.8  | Q40 | 14,417 | 21.5  |
| Q8  | 16,807 | 9.6  | Q24 | 14,910 | 8.0  | Q41 | 14,360 | 30.1  |
| Q10 | 16,756 | 10.4 | Q25 | 14,897 | 7.9  | Q42 | 14,221 | 40.5  |
| Q11 | 16,730 | 15.7 | Q28 | 14,827 | 10.7 | Q43 | 14,123 | 48.2  |
| Q12 | 16,685 | 14.7 | Q29 | 14,811 | 12.0 | Q44 | 13,885 | 88.5  |
| Q15 | 16,620 | 23.4 | Q30 | 14,785 | 8.9  | Q45 | 13,775 | 132.8 |
| Q16 | 16,576 | 11.5 | Q31 | 14,777 | 8.9  | Q46 | 13,659 | 125.3 |
| Q18 | 16,497 | 33.8 | Q33 | 14,707 | 20.0 | | | |
| Q20 | 16,399 | 34.7 | Q35 | 14,667 | 32.5 | | | |
|     |        |      | Q36 | 14,626 | 12.2 | | | |
|     |        |      | Q38 | 14,567 | 20.3 | | | |

**Table-IV:** List of the parameters obtained from fitting temperature-dependence of the prominent intra-configurational transition modes of Nd$^{3+}$ in Nd$_2$ZnIrO$_6$ via using equations as described in the text. Units are in cm$^{-1}$, $W_i$ is a dimensionless parameter.

| Mode | $\omega(0)$ | $\alpha_i$ | $W_i$ |
|---|---|---|---|
| Q2  | 17042.4 | -102.1 | -0.10 |
| Q5  | 16926.3 | -90.4  | -0.09 |
| Q6  | 16875.5 | -87.3  | -0.08 |
| Q10 | 16755.9 | -31.5  | -0.03 |
| Q11 | 16730.4 | -51.1  | -0.05 |
| Q15 | 16620.6 | -106.3 | -0.11 |
| Q16 | 16575.8 | -50.6  | -0.05 |
| Q22 | 14989.7 | -72.8  | -0.07 |
| Q23 | 14938.5 | -58.9  | -0.06 |
| Q29 | 14810.1 | +57.0  | +0.06 |

**FIGURE CAPTION:**

**FIGURE 1:** (Color online) (a) Raman spectra of $Nd_2ZnIrO_6$ excited with 633 nm. The solid thick red line indicates a total fit to the experimental data and the solid thin blue line corresponds to individual peak fitting, label P1-P23 are phonon modes. Inset (a1) shows the temperature evolution of the Raman spectra in the temperature range of 4-330 K excited with 633 nm, and (a2) shows the comparison of Raman spectra excited with 633 nm and 532 nm laser at 4 K and 330 K and label R1-R7 corresponds to crystal-field modes.

**FIGURE 2:** (Color online) (a) Calculated eigen vectors of the phonon modes P1 to P23, where green, blue, pink and red spheres represents the Nd, Ir, Zn and Oxygen atoms, respectively. Black arrows on the atoms indicates the direction of atomic displacement and the arrow size illustrate the magnitude of atomic vibration. a, b and c indicate the crystallographic axis.

**FIGURE 3:** (Color online) Temperature-dependence of the mode frequencies and line-widths of the first-order phonon modes P1, P3-P4, P7-P8, P10-P11, P15 and P20-P23. Blue and red spheres represent experimental value of frequencies and line-widths, respectively. Solid red and blue lines in temperature range are the fitted curves as described in the text. Solid black lines are guide to the eye and broken lines are extended estimation of fitted curves. Shaded portion below ~ 100 K depicts the region where pronounced phonon anomalies are observed.

**FIGURE 4:** (Color online) Angular-dependence of the Raman spectrum of $Nd_2ZnIrO_6$; (a and b) shows the raw data at different incident light angle in range of 0º to 90º (inset show from 90º to 180º) in the spectral range of 50-330 $cm^{-1}$ and 330-800 $cm^{-1}$, respectively. (c) Left side shows the optical micrograph of the microcrystal and right side shows the schematic depicting the direction of polarization of incident and scattered light in the *ab*-plane. $\theta_o$ is the angle between scattered light polarization vector ($\hat{e}_s$) and crystal *a*-axis, and $\theta$ is the angle between $\hat{e}_i$ and $\hat{e}_s$. $k_i$ and $k_s$

represents the direction of incident and scattered beam. (d) Polar plots of the angular-dependence intensity of the prominent phonon modes. The solid curves are the fitted with angular-dependence mode intensity equations as described in text.

**FIGURE 5:** (Color online) (a) Temperature-dependence of the frequency and line-width of the prominent crystal-field modes R2, R4 and R7; inset of top panel shows frequency difference ( $\Delta\omega_{21} = \omega_2 - \omega_1$ ) between mode R2 and R1 as function of temperature. (b) Temperature-dependence of the intensity of modes R2, R4 and R7 along with those of first-order phonon modes P1, P3 and P4. Inset in top and middle panel shows the change in slope of the intensity near magnetic transition temperature, solid lines are guide to the eye; inset in bottom panel shows the intensity ratio (i.e. R2/R1) and solid line is the fitted curve as described in the text. (c) Temperature evolution of the modes (R1-R7) in the spectral range of 20-500 cm$^{-1}$ excited with 532 nm laser. Inset shows the polarization-dependence of the modes R1, R3, R5 and R7 at 4 K, and schematic for the CF splits levels of the Nd$^{3+}$ ground state ($^4I_{9/2}$) multiplets into five Kramer's doublets. (d) Shows the polarization-dependence of R$_2$, R$_4$ and R$_6$ modes showing quasi-isotropic nature.

**FIGURE 6:** (Color online) (a and b) Raman spectra of Nd$_2$ZnIrO$_6$ in the frequency range of 1700-4300 cm$^{-1}$ and 4300-5500 cm$^{-1}$, respectively, recorded at 75 K with 532 nm laser excitation and label Q1-Q46 corresponds to intra-configurational modes (i.e. electronic transition modes of 4*f*-levels of Nd$^{3+}$). The spectra are fitted with a sum of Lorentzian functions, where the solid thick red line indicates a total fit to the experimental data and the solid thin blue line corresponds to individual peak fit. The shaded region shows the same spectra in absolute energy scale. (c) Schematic of the CF levels of Nd$^{3+}$ ion and vertical blue lines depicts the observed transition.

**FIGURE 7:** (Color online) (a and b) Temperature-dependence of frequency and line-width of intra-configurational modes. Solid red lines are fitted curves as described in text and solid black

and blue lines are guide to the eye. (c) Normalized intensity of the prominent modes Q2, Q5-Q7, Q10-Q11, Q15-Q16, Q22-Q23, and Q28-Q29.

**Figure 1:**

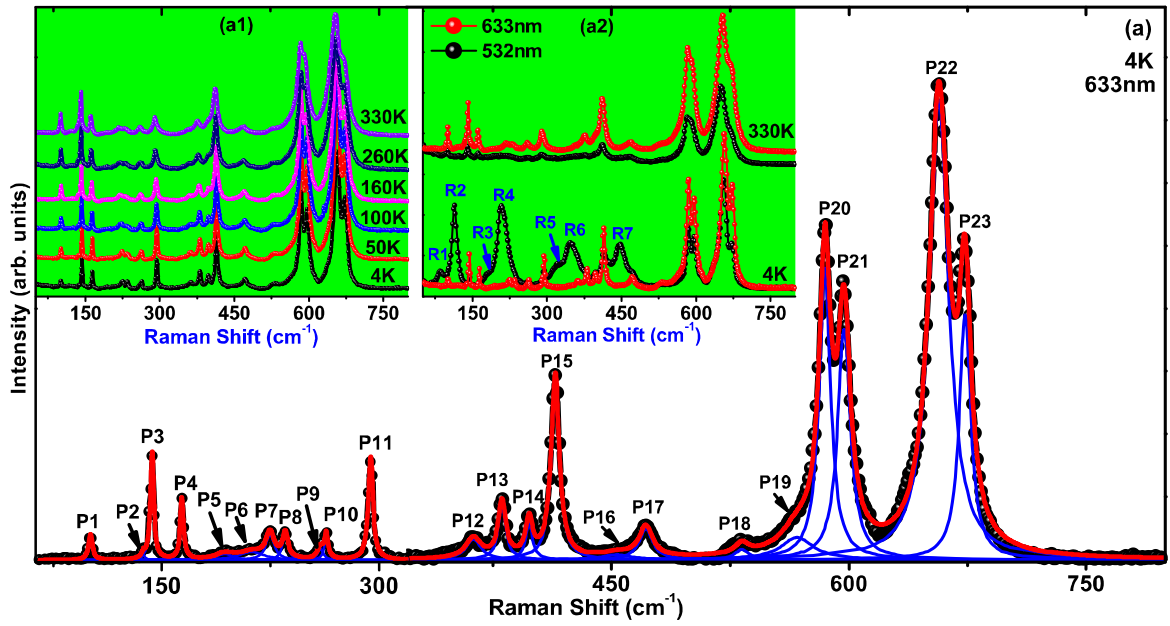

**Figure 2:**

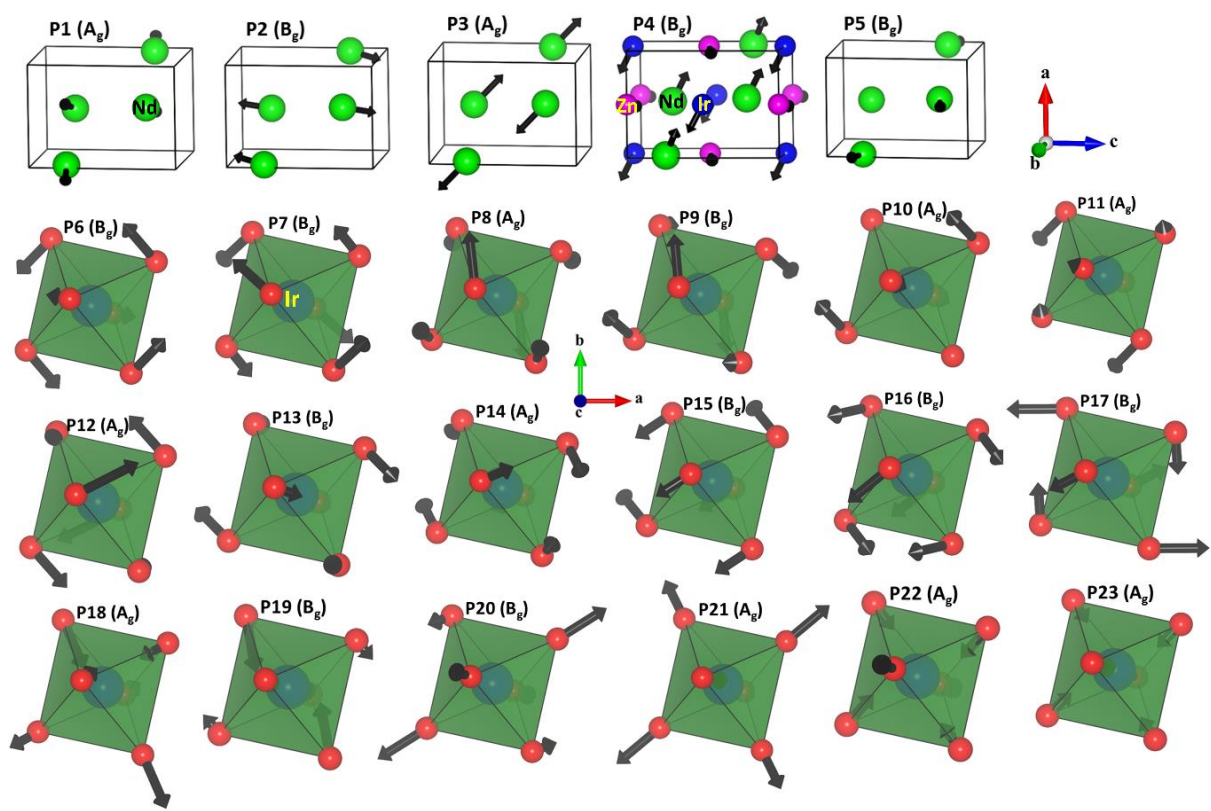

**Figure 3:**

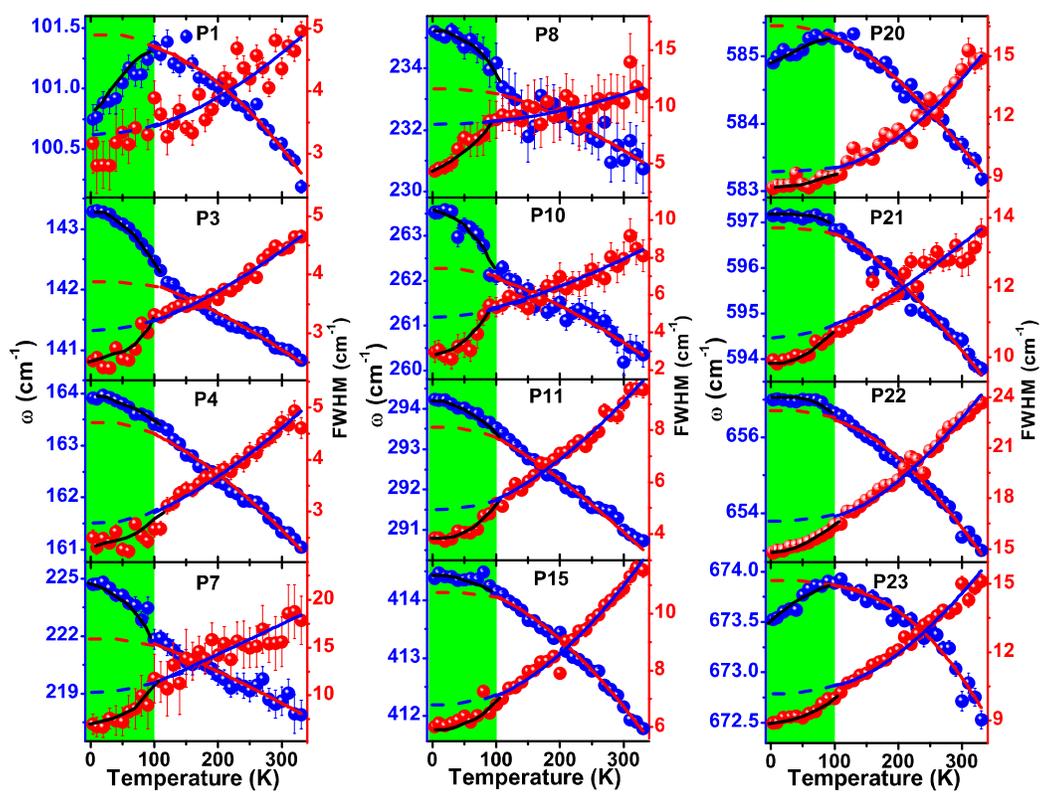

**Figure 4:**

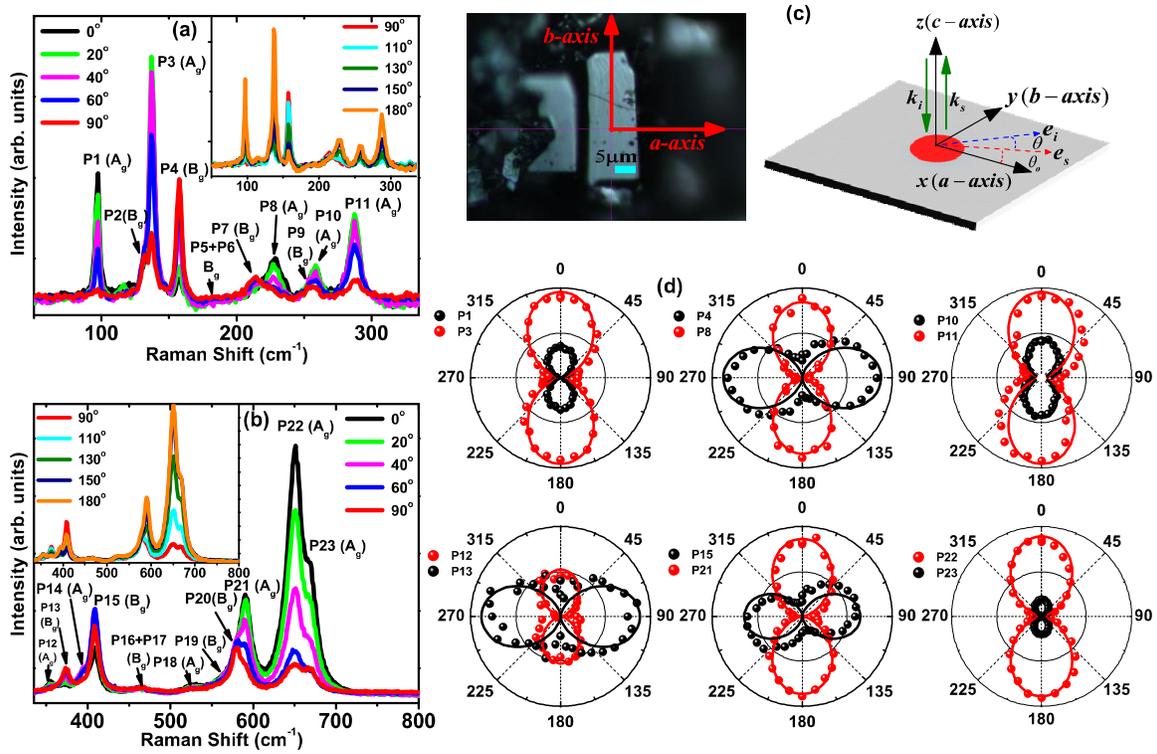

**Figure 5:**

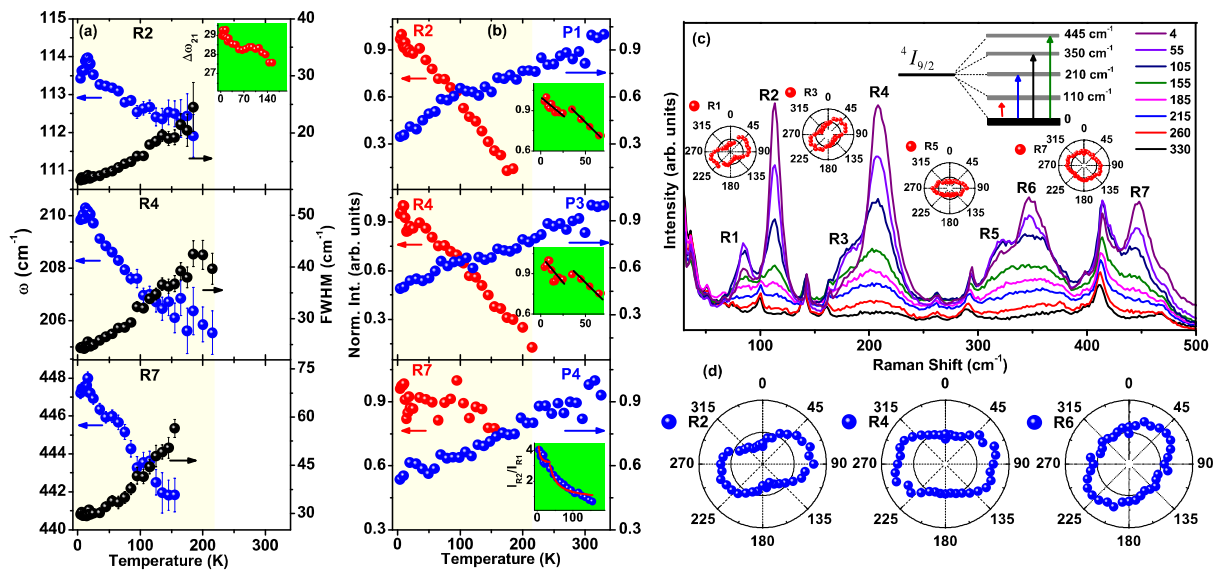

**Figure 6:**

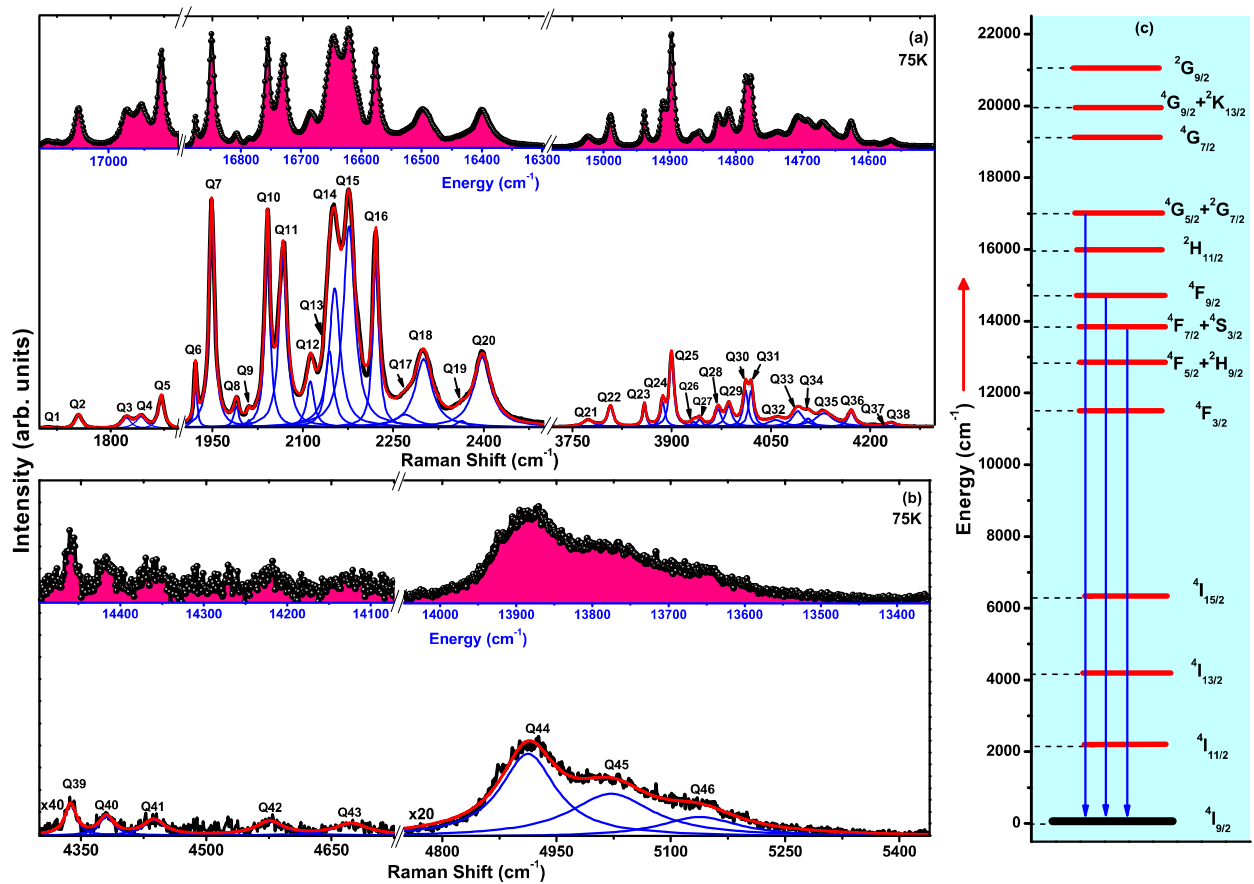

**Figure 7:**

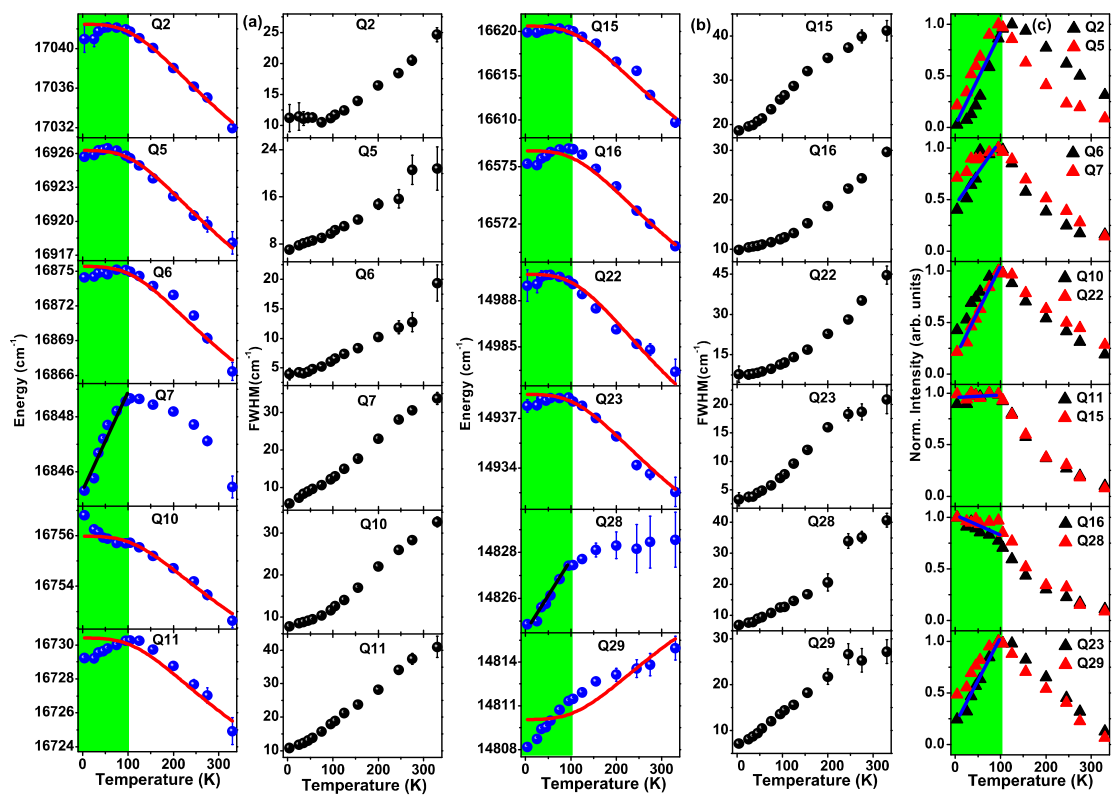